\documentclass[letterpaper]{article} 
\usepackage{aaai24}  
\usepackage{times}  
\usepackage{helvet}  
\usepackage{courier}  
\usepackage[hyphens]{url}  
\usepackage{graphicx} 
\usepackage{natbib}  
\usepackage{caption} 
\usepackage{booktabs} 
\usepackage{enumitem}
\usepackage{multirow}
\usepackage{amsmath}
\usepackage{amsthm}
\usepackage{amssymb}
\usepackage{cleveref}
\usepackage{dsfont}
\usepackage{soul}
\usepackage{makecell}

\usepackage{xspace}

\newcommand{\bx}{{\boldsymbol x}\xspace}
\newcommand{\bX}{{\boldsymbol X}\xspace}
\newcommand{\bz}{{\boldsymbol z}\xspace}

\usepackage[usenames, dvipsnames]{xcolor}

\usepackage{xargs}
\usepackage[colorinlistoftodos,prependcaption,textsize=tiny]{todonotes}

\title{Calibrate-Extrapolate: Rethinking Prevalence Estimation with Black Box Classifiers} 

\author{
Siqi Wu,
Paul Resnick\\
}

\affiliations {
University of Michigan, Ann Arbor\\
\{siqiwu, presnick\}@umich.edu
}

\begin{document}
\maketitle

\begin{abstract}

In computational social science, researchers often use a pre-trained, black box classifier to estimate the frequency of each class in unlabeled datasets. A variety of prevalence estimation techniques have been developed in the literature, each yielding an unbiased estimate if certain stability assumption holds. This work introduces a framework to rethink the prevalence estimation process as calibrating the classifier outputs against ground truth labels to obtain the joint distribution of a base dataset and then extrapolating to the joint distribution of a target dataset. We call this framework ``Calibrate-Extrapolate''. It clarifies what stability assumptions must hold for a prevalence estimation technique to yield accurate estimates. In the calibration phase, the techniques assume only a stable calibration curve between a calibration dataset and the full base dataset. This allows for the classifier outputs to be used for disproportionate random sampling, thus improving the efficiency of calibration. In the extrapolation phase, some techniques assume a stable calibration curve while some assume stable class-conditional densities. We discuss the stability assumptions from a causal perspective. By specifying base and target joint distributions, we can generate simulated datasets, as a way to build intuitions about the impacts of assumption violations. This also leads to a better understanding of how the classifier's predictive power affects the accuracy of prevalence estimates: the greater the predictive power, the lower the sensitivity to violations of stability assumptions in the extrapolation phase. We illustrate the framework with an application that estimates the prevalence of toxic comments on news topics over time on Reddit, Twitter/X, and YouTube, using Jigsaw's Perspective API as a black box classifier. Finally, we summarize several practical advice for prevalence estimation.

\end{abstract}

\section{Introduction}
\label{sec:intro}

Measuring the frequency of certain labels within a data sample is a common task in many disciplines. This problem, generally called ``prevalence estimation'' or ``quantification'', has a wide range of real world applications, from quantifying the number of infected COVID-19 patients in a country~\cite{sempos2021adjusting}, to automated accounts in a social platform~\cite{yang2020scalable}, and to anti-social posts in an online community~\cite{park2022measuring}. It would be ideal to manually examine every individual in the dataset. This approach is called ``census''. However, census is often prohibitively expensive. 

In computational social science (CSS), it is common to make use of an imperfect pre-trained classifier that assigns a label, or label probability, to each item~\cite{yang2020scalable,rajadesingan2021political,saha2023rise,pfeffer2023just}. The prevalence estimation techniques then somehow account for the imperfections of the classifier. Alternative approaches are possible, such as a model trained with prevalence estimation evaluation metrics rather than individual classification accuracy as the objective function~\cite{esuli2015optimizing}. However, the classifier-aided approaches are especially attractive when the dataset is on a large scale, ground truth labels are difficult or expensive to obtain, and a high quality pre-trained classifier such as ChatGPT or Jigsaw's Perspective API \cite{wulczyn2017ex} is available for use.

Extant literature on prevalence estimation with black box classifiers is plentiful and there are two useful survey articles \cite{forman2008quantifying,gonzalez2017review}. In this paper, we do not propose any new estimation techniques. Instead, our contribution is a conceptual framework called ``Calibrate-Extrapolate''. It rethinks the prevalence estimation process not as merely estimating the percentage of ground truth labels but as estimating the complete joint distribution between the classifier outputs and ground truth. A calibration phase gathers ground truth labels for a small data sample that is purposefully selected from a base dataset, and then estimates the joint distribution for the full base dataset. An extrapolation phase assumes that some properties of the base joint distribution are stable and infers other properties from the observed classifier output distribution of a target dataset. In both phases, the prevalence for label of interest can be read off from the inferred joint distribution of classifier outputs and ground truth labels. 

The Calibrate-Extrapolate framework is broadly applicable to many real world settings. It is also flexible because researchers can still customize design elements in some steps. We identify four such elements: which black box classifier to apply, which data to sample for ground truth labels, which stability assumption to make, and which prevalence estimation technique to employ. 

In a real world setting, it may be unclear which stability assumptions are reasonable to make. Thinking about prevalence estimation techniques in the Calibrate-Extrapolate framework makes it clear what assumptions each technique requires about the stable properties (or lack thereof) between training, base, and target datasets; and how violations of those assumptions can lead to errors in the final prevalence estimate. For example, assuming only stable calibration curve between base and target datasets limits the range of possible final estimates and attenuates any observed change in the density of classifier outputs. On the other hand, assuming only class-conditional densities amplifies any observed change in the density of classifier outputs and can lead to estimates clipped at 0 or 100\%. In addition, thinking about prevalence estimation in this framework clarifies the value of a more accurate classifier. While a weak classifier can still yield correct prevalence estimates if repeated many times~\cite{teblunthuis2024misclassification}, those estimates will be less certain and less robust to incorrect stability assumptions between base and target datasets.

We demonstrate the impacts of alternative design elements using simulated data, where large-scale base and target datasets are generated from known joint distributions of classifier outputs with ground truth labels. We experimented with two kinds of data generating processes (intrinsic and extrinsic), two classifiers (strong and weak), two stability assumptions (stable class-conditional densities and stable calibration curve), and two prevalence estimation techniques (mixture model and probabilistic estimator). These yield an instructive set of comparisons that highlight both the strengths and limitations of each estimation technique. 

We apply the Calibrate-Extrapolate framework to estimating the weekly prevalence of toxic comments on news topics on Reddit, Twitter/X, and YouTube in 2022. We selected Jigsaw's Perspective API as our black box classifier and sent all collected comments to it for obtaining a classifier score for each. We used comments from August 2021 as the base dataset, asked crowdworkers to annotate it, and used the annotations to calibrate the Perspective API outputs. We then extrapolated the learned base joint distribution to make an inference for target datasets in later weeks. The alternative prevalence estimation techniques yield a range of estimates for each platform for each week. We estimated consistently higher prevalence of toxic comments on YouTube (12.93\%) than on Twitter/X (9.39\%), than on Reddit (7.75\%). 

In sum, the contributions of this work include:

\begin{itemize}
    \item a framework that rethinks the prevalence estimation process as modeling the joint distribution between classifier outputs and ground truth labels in a dataset;
    \item experiments and arguments about how different design elements of the prevalence estimation process impact the accuracy of estimates, including guidance about avoiding clear missteps and ways to think about robustness to incorrect stability assumptions;
    \item repeated empirical prevalence estimates, over time, of toxic comments on news topics on Reddit, Twitter/X, and YouTube.
\end{itemize}

\section{Background}
\label{sec:background}

For concreteness and simplicity, assume there are only two ground truth classes, which we will call positive $\oplus$ and negative $\ominus$, and positive is the class of interest. We are interested in prevalence estimation processes that make use of a black box classifier. It is black box in the sense that it was previously trained by other people on different datasets, and the prevalence estimator cannot look inside the classifier to change how it works. But we can apply the classifier to new data. Formally, given the feature vector $\bx_i$ for an item $i$, a classifier $C(\cdot)$ would output a continuous score $C(\bx_i) \in [0, 1]$. A few widely used black box classifiers in the CSS community include Perspective API for toxicity detection~\cite{wulczyn2017ex}, VADER for sentiment analysis~\cite{hutto2014vader}, Botometer for online bot detection~\cite{yang2020scalable}, and LLMs for making inferences in a variety of text classification tasks.

A naive approach of prevalence estimation would just count the number of items with classifier scores above some threshold, or sum up the classifier scores as if they were probabilities. However, neither would yield very good prevalence estimates for two reasons. 

First is calibration. When a classifier outputs a score between 0 and 1, while a high value indicates higher confidence that the ground truth label is positive, it is not clear that the score is intended to be interpreted as a {\it calibrated} probability. It is thus not safe to assume that, given a collection of items all receiving a classifier score of 0.8, on average 80\% of items would have positive ground truth labels. For example, prior work finds that deep neural networks tend to generate overconfident outputs. Several methods have been developed to calibrate the classifier outputs to true probabilities~\cite{platt1999probabilistic,guo2017calibration}. 

Second is dataset shift. The classifier may have been trained on a dataset that is systematically different from the dataset for which the researchers are estimating prevalence. For example, the Perspective API was trained on comments from forums such as Wikipedia and the New York Times. However, researchers have applied Perspective API to social media comments from Reddit~\cite{rajadesingan2021political}, Twitter/X~\cite{hua2020towards}, and YouTube~\cite{wu2021cross}. The linguistic features of those comments may differ in important ways from the training data. Furthermore, even if the classifier is trained on a similar data source, it is possible that new terms might appear between the training and inference time, resulting in more frequent errors for those new terms. Dataset shift is a very important problem in the machine learning literature~\cite{moreno2012unifying,arora2021types} and especially a concern for lexicon and rule-based classifiers such as VADER.

\subsubsection{Prevalence estimation techniques}
In a recent survey of prevalence estimation, \citet{gonzalez2017review} categorized estimation techniques into three groups: classify count \& correct, distribution matching, and quantification learning. We will review the first two groups and rethink them in terms of the Calibrate-Extrapolate framework in~\Cref{sec:framework}. The quantification learning methods are out of scope for this paper because they directly optimize a loss function for prevalence estimation evaluation metrics, which is akin to training a completely new model rather than post-processing outputs from a pre-trained classifier. We remark that the idea of classifier posterior calibration, which is what we do in the calibration phase of the Calibrate-Extrapolate framework, is similar to fine-tuning a pre-trained model to specific datasets and to calibrated probabilities~\cite{guo2017calibration}. 

\subsubsection{Linking stability assumptions to data generating processes}
Current research in machine learning has offered a causal perspective for thinking about potential distribution shifts across datasets. This perspective considers the causal relation between a vector of covariates $\bx$ and the labels $y$~\cite{scholkopf2012causal,jin2021causal}. Thinking of either the covariates causing labels, or the labels causing the covariates, leads to two different kinds of data generating processes and thus two different distribution shifts: 

\begin{itemize}
    \item \textit{Intrinsic data generation $\rightarrow$ Prior probability shift}. 
    In one conceptual model of the data generating processes, sometimes called ``intrinsic data generation''~\cite{card2018importance}, the label $y$ causally determines the covariate $\bx$ (causal chain: $y \rightarrow \bx$). That is, we assume that $y_i$ is drawn from some distribution and the realized $y_i$ decides the covariates $\bx_i$. For example, when simulating, we might generate $\bx$ as independent draws from different distributions for the positive or negative class. This would lead to a stable property across a base dataset $B$ and a target dataset $T$: $P_B(\bx|y) = P_T(\bx|y)$. While that property is stable, there can be what is called prior distribution shift or label shift: $P_B(y) \neq P_T(y)$~\cite{moreno2012unifying,scholkopf2012causal}. A real world example is detecting if buyers recommend a product in online reviews: they start with an intrinsic opinion $y$, which is either {\it recommend} or {\it not recommend}, and then express the opinion in words $\bx$.
    \item \textit{Extrinsic data generation $\rightarrow$ Covariate shift}. 
    With extrinsic data generation, the covariate $\bx$ causally determines the ground truth label $y$ (causal chain: $\bx \rightarrow y$). That is, we assume that $\bx_i$ is drawn from some distribution and the realized $\bx_i$ decides the label $y_i$. This would lead to a different stable property across a base dataset $B$ and a target dataset $T$: $P_B(y|\bx) = P_T(y|\bx)$. While that property is stable, there can be a covariate shift: $P_B(\bx) \neq P_T(\bx)$. A real world example is detecting if other users find an online product review helpful: they evaluate the review body based on extrinsic features $\bx$ such as review {\it length}, if {\it has\_images}, and/or if {\it is\_polite}, and then make a judgment $y$, which is either {\it helpful} or {\it not helpful}.
\end{itemize}

For this paper, we insert one more element into the causal analysis: the classifier output $C(\bx)$. The classifier makes predictions deterministically based on the feature vectors $\bx$. Thus the causal chain is always $\bx \rightarrow C(\bx)$. Our framework aims to estimate the joint distribution $P(C(\bx), y)$. We are more interested in what properties stay the same across datasets, rather than what may have shifted, because stable properties play a key role in the extrapolation phase of our framework. To this end, we reframe the terminology in distribution shift research to stability assumptions.

\begin{itemize}
    \item \textit{Prior probability shift $\rightarrow$ Stable class-conditional densities}. 
    Given intrinsic data generation and prior probability shift, we can derive that $y$ also causally determines the classifier output $C(\bx)$. By the law of total probability: 
    $$ P(C(\bx)|y) = \sum\limits_{\bz \in \bX} P(C(\bx)|\bz,y) * P(\bz|y) $$
    Because the classifier's output is deterministic and depends only on the feature input $\bz$, but not on $y$. In other words, $y$ does not provide additional information about the classifier output given $\bz$:
    $$ P(C(\bx)|\bz,y) = P(C(\bx)|\bz) $$
    Therefore, 
    $$ P(C(\bx)|y) = \sum\limits_{\bz \in \bX} P(C(\bx)|\bz) * P(\bz|y) $$ 
    $P(C(\bx)|\bz)$ is stable because the classifier is unchanged between datasets $B$ and $T$.
    $P(\bz|y)$ is stable under the intrinsic data generating process.
    Thus, class-conditional densities $P(C(\bx)|y)$ are also stable between datasets.
    $$ P_B(C(\bx)|y) = P_T(C(\bx)|y) $$ 
    
    \item \textit{Covariate shift $\nrightarrow$ Stable calibration curve}. 
    Given extrinsic data generation and covariate shift, we might expect that we can infer the analogous stability property, $P_B(y|C(\bx)) = P_T(y|C(\bx))$. However, this is not quite true. By the law of total probability, and the fact that feature input $\bz$ causally determines $y$:
    \begin{equation*}
    \begin{split}
    P(y|C(\bx)) & = 
    \sum\limits_{\bz \in \bX} P(y|\bz,C(\bx)) * P(\bz|C(\bx)) \\
    & = \sum\limits_{\bz \in \bX} P(y|\bz) * P(\bz|C(\bx)) \\
    \end{split}
    \end{equation*}
    For any $\bz$, $P(y|\bz)$ is stable given the extrinsic data generating process. However, we could have $ P_B(\bz|C(\bx)) \neq P_T(\bz|C(\bx))$.  For example, imagine only two possible feature vectors, $\bz_1$ and $\bz_2$, both yield a particular classifier score. But suppose that $\bz_1$ is much more common in dataset $B$ while $\bz_2$ is more common in dataset $T$. In that case, $P(\bz_1|C(\bx))$ decreases and $P(\bz_2|C(\bx))$ increases from $B$ to $T$. Therefore, calibration curve $P(y|C(\bx))$ are not stable between datasets.\footnote{Stability of the calibration curve will hold under some stricter assumptions. For example, if for each classifier output value $C(\bx)$, there is one and only one feature vector $\bz$ corresponding to it. See more discussion in~\cite{tasche2022class}.}
    $$ P_B(y|C(\bx)) \neq P_T(y|C(\bx)) $$
    If a prevalence estimation technique (e.g., probabilistic estimator) depends on the calibration curve $P(y|C(\bx))$ being stable, this will require an additional assumption beyond just extrinsic data generation. We refer to this as the stable calibration curve assumption.  
\end{itemize}

\section{The Calibrate-Extrapolate Framework}
\label{sec:framework}

\begin{figure*}[htbp]
    \centering
    \includegraphics[width=1\linewidth]{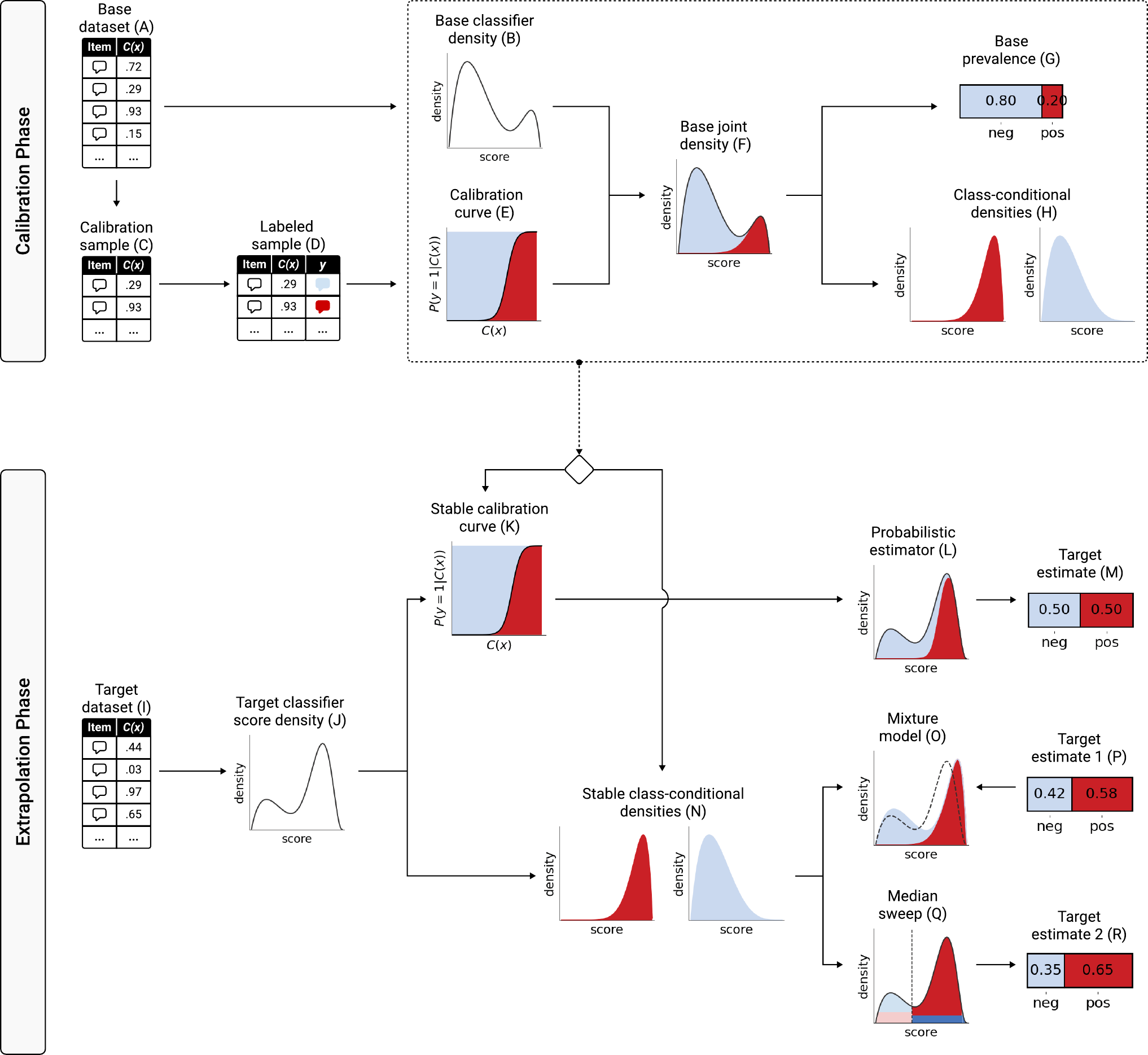}\\
    \caption{
    The Calibrate-Extrapolate framework. 
    (top panel) The calibration phase aims to estimate the label prevalence for a base dataset $B$. We apply a black box classifier $C(\cdot)$ to the feature vector $\bx$ of each item in $B$ and obtain a base classifier score density (step A $\rightarrow$ B). We curate a calibration sample $S$ from $B$ (step A $\rightarrow$ C). We then obtain the ground truth labels for $S$ (step C $\rightarrow$ D). Next, we fit a calibration curve function on $S$ (step D $\rightarrow$ E). We can use the calibration curve of $S$ and the observed classifier score density of $B$ to compute the joint distribution between $C(\bx)$ and $y$ for $B$ (step B + E $\rightarrow$ F). This is assuming calibration curves between $B$ and $S$ are stable, i.e., $P_S(y|C(\bx)) = P_B(y|C(\bx))$. From the joint distribution, we can derive the class-conditional densities for positive and negative class in $B$ (step H) and make a prevalence estimate of $B$ (step G).
    (bottom panel) The extrapolation phase aims to estimate the label prevalence in a target dataset $T$. We apply the same classifier $C(\cdot)$ to all items in $T$ (step I $\rightarrow$ J). By making a stability assumption (step K, N) that links the joint distributions of $B$ and $T$, we can infer the joint distribution of $T$ (step L, O, Q) and read off the final prevalence estimate of $T$ (step M, P, R).
    }
    \label{fig:prevalence_framework}
\end{figure*}

\Cref{fig:prevalence_framework} illustrates the Calibrate-Extrapolate prevalence estimation framework. To present alternative prevalence estimation processes within the framework, we use a running example, estimating the fraction of people registered as members of the Republican Party among users of a U.S. social media platform at two different points in time, a base period and a target period. Assume that we have a pre-trained political leaning classifier that takes as input a corpus of text written by a person and outputs a $1$ if it is sure the writer is a Republican, $0$ if sure they are not, and numbers in between to express varying degrees of confidence. This yields a base dataset of people with corresponding classifier scores $C(\bx)$ (A in \Cref{fig:prevalence_framework}). Those classifier scores can be plotted as an empirical classifier score density curve (B in \Cref{fig:prevalence_framework}).

\subsection{Calibration Phase}
\label{ssec:calibration_phase}

In the calibration phase, we select a limited sample for which we will acquire ground truth labels $y$ based on their voter registration records (C and D in \Cref{fig:prevalence_framework}). Following the current conventional color-coding in U.S. politics, we use red for Republicans and blue for Democrats. Getting the labels may be an expensive process, requiring verifying real names and accessing records from local municipalities; hence, we carry it out only for a limited sample. The sample may be selected at random. Alternatively, to make the best use of a limited set of expensive ground truth labels, it can be helpful to use the classifier to select a disproportionate random sample based on the classifier outputs. For example, it may be helpful to oversample in ranges of $C(\bx)$ that occur rarely, especially for highly imbalanced datasets and that imbalance reflects on the distribution of $C(\bx)$. Two sampling strategies can possibly improve the efficiency of final base dataset prevalence estimate (i.e., have a smaller confidence interval around it): one is to sample equally from every $C(\bx)$; the other is called the ``Neyman allocation''~\cite{neyman1934two}.

From the labeled calibration sample, we can estimate a calibration curve, which is a function that maps the classifier outputs to calibrated probabilities (E in \Cref{fig:prevalence_framework}). There are several options for this estimation process. One is non-parametric binning: divide the $C(\bx)$ values into bins; compute the empirical fraction of positives (Republicans) in each bin, and treat that fraction as the calibrated probability for any $C(\bx)$ within the bin. Isotonic regression adds an extra constraint that calibrated probabilities must be higher for larger $C(\bx)$ values~\cite{chakravarti1989isotonic}. Platt scaling makes a further parametric assumption of a smooth sigmoid curve, as we have illustrated in the figure; it is estimated through a logistic regression that treats $C(\bx)$ as the independent variable and the binary ground truth $y$ as the dependent variable~\cite{platt1999probabilistic}. Finally, temperature scaling extends Platt scaling by adding one more parameter to control the scaling strength~\cite{guo2017calibration}.

Note that a classifier density function (B in \Cref{fig:prevalence_framework}) and a classifier calibration curve (E in \Cref{fig:prevalence_framework}) together define the joint distribution of classifier scores and ground truth labels. To see that, imagine using the calibration curve to color each vertical slice of the classifier density plot, yielding the more familiar representation of a joint distribution as a stacked joint density plot (F in \Cref{fig:prevalence_framework}). At low $C(\bx)$ values, the vertical slice is colored almost entirely blue; while at high values, almost entirely red.

Alternatively, three other properties can be used to define a joint distribution: the ground truth prevalence (G in \Cref{fig:prevalence_framework}) and two class-conditional density functions: one for Republicans and one for Democrats (H in \Cref{fig:prevalence_framework}). Given a representation of the joint distribution, it is possible to read off all these properties. For example, given the joint density plot (F), the blue and red areas under the curve indicate the ground truth prevalence (G), and looking at the blue and red areas separately from the stacked joint density plot yields density plots for each class (H). Note that (G) directly estimates the prevalence of Republicans in the base dataset, the quantity of interest. However, other properties of the base joint distribution will be important for the extrapolation phase, it is thus helpful to think about the calibration phase as creating an estimate of the complete joint distribution.

Before turning to the extrapolation phase, it is worth reinterpreting some existing prevalence estimation techniques in terms of processes yielding the base prevalence estimate in step G of \Cref{fig:prevalence_framework}. One is \textit{calibrated probabilistic classify and count} \cite{bella2010quantification,card2018importance}. The calibrated part refers to passing each item's classifier score through the calibration curve (E) to yield a calibrated probability. The sum of those probabilities is divided by the total number of items, yielding the prevalence estimate. The summation of probabilities is equivalent to first counting how many items have each classifier score and then multiplying the corresponding calibrated probabilities by the frequency of classifier scores; that is what happens when we integrate across the possible density scores in (F) to compute the size of the red area (G). 

An even simpler version is an uncalibrated \textit{probabilistic classify and count}. This technique skips the steps of sampling and labeling (B and C in \Cref{fig:prevalence_framework}) entirely and just adds up the classifier scores as if they were calibrated probabilities. This is equivalent to assuming, without estimation,  that the calibration curve (E) is a straight line from $(0,0)$ to $(1,1)$: $P(y=\oplus|C(\bx)) = C(\bx)$. If the classifier is indeed perfectly calibrated, the prevalence estimate will be fine; but if not, the estimate could be far off.

Other approaches binarize the classifier outputs by applying a threshold. For example, any classifier scores above 0.7 might be treated as positive and lower scores as negative. Note that the joint distribution becomes just a confusion matrix when applying such a threshold. This leads to the \textit{classify and count} and \textit{adjusted classify and count} techniques. The unadjusted version simply treats the fraction of items with $C(\bx)$ scores above the threshold as the prevalence estimate. The adjusted version computes a multiplier that is applied to the estimate. In our framework, these approaches are equivalent to collapsing the calibration curve (E) into a step function, with one universal probability for scores above the threshold and another for scores below the threshold. In the unadjusted version, one assumes perfect classification: for scores above the threshold the calibrated probability is 1 and for those below it is 0. In the adjusted version, the empirical calibration curve is computed and then it is collapsed to provide two scores; these scores correspond to the multipliers in the adjusted classify and count technique. 

The framework visually reveals that applying thresholds in this way is risky. The step-function calibration curve treats all items with classifier scores on the same side of the threshold as if they have the same probability of a ground truth positive label. But an item with a score just above the threshold may be less likely to be ground truth positive than one with a classifier score of almost 1. If the frequency of items near and far from the threshold varies, the accuracy of the calibrated estimate will also vary. 

Another point of caution is that it is not safe to estimate class-conditional densities for the base dataset directly from the labeled calibration sample (D) because disproportionate sampling alters the observed class-conditional densities. For example, imagine a scenario where items of $C(\bx) \in [0.9, 1]$ are rare, and we oversample those. If items of $C(\bx) \in [0.9, 1]$ are more likely to be ground truth positive, as we would expect, this will artificially increase the density in $[0.9, 1]$ for the positive class density function. Instead of calculating the class-conditional densities directly from the calibration sample, the correct approach is to first infer the calibration curve (E), then, together with the base classifier density, recover the full joint distribution, from which the class-conditional densities can be inferred. 

It is also worth noting that if the goal is only to produce a single prevalence estimate for a base dataset, a random sample would yield an unbiased estimate. Indeed, the final base prevalence estimate from the calibration phase (G) is expected to the same as one would get by simply taking the fraction of positive items in a randomly selected calibration sample. However, the classifier helps in making a prevalence estimate for the base dataset by improving efficiency, allowing the same number of labels to yield a tighter confidence intervals around the base prevalence estimate. This is achieved through purposefully oversampling minority classes indicated by the classifier scores.

\subsection{Extrapolation Phase}
\label{ssec:extrapolation_phase}

As we shall see, the real power of having a black box classifier appears when we want to reuse it to make prevalence estimates for additional target datasets. Having estimated the joint distribution of classifier output and ground truth on the base dataset in the calibration phase, in the extrapolation phase we can make some assumptions about properties of the base joint distribution that would remain stable. Two lines of estimation techniques make different stability assumptions (K and N in \Cref{fig:prevalence_framework}). 

First, consider the left-side representation of a joint distribution as defined by a classifier density function (B) and a calibration curve (E). We assume that the classifier density function from the base joint distribution is stable and thus also applies to the target dataset (K). The classifier density function, however, may have changed (e.g., J differs from B). This intuitively corresponds to an extrinsic data generating process where the covariates $\bx$ causally determine the ground truth label $y$. But as we noted in~\Cref{sec:background}, the assumption of a stable calibration curve requires an additional assumption beyond just extrinsic data generation. 

This leads to the \textit{probabilistic estimator} technique~\cite{bella2010quantification}. The target classifier density function is estimated from the empirical frequency of classifier scores in a target dataset (J). Together with the calibration curve (K) borrowed from the base dataset, that fully determines the target joint distribution. The joint density graph, with colorings (L), can be derived by coloring vertical slices of J using K. The prevalence estimate for the target dataset (M) can be read off from the joint density by integration, computing the size of the blue and red areas under the curve.

Next, consider the right-side representation of the base joint distribution as defined by a ground truth prevalence (G) and the class conditional densities (H). Here, we assume that the class conditional densities are stable (N). The ground truth prevalence, however, may have changed.  As noted in \Cref{sec:background}, this intuitively corresponds to an intrinsic data generation process where the ground truth label $y$ (e.g., being a Republican) causally determines the observed features $\bx$ (e.g., the text that was written), which causally determines the resulting classifier score $C(\bx)$ (e.g., a classifier judges it is written by a Republican).

Two techniques are based on this second stability assumption. One is called the \textit{mixture model}~\cite{forman2005counting}. It treats the observed frequency of classifier scores for the target dataset as a hint about the true label prevalence. It does a grid search over the range of possible prevalence estimates. Each possibility defines a mixture of the two class-conditional densities and thus generates an implied classifier score density function. A distance metric needs to be introduced to compare the implied classifier density to the observed classifier score density (J in \Cref{fig:prevalence_framework}). Several distance metrics are possible; one popular choice in the literature is Hellinger distance~\cite{gonzalez2013class}. The final target prevalence estimate (P) is the value that minimizes the selected distance function. Note that, the optimal estimate may still yield an implied classifier score density not perfectly matching (J), illustrated as the gap between the dashed line and filled area in (O).

The other technique that assumes stable class-conditional densities is called \textit{median sweep}~\cite{forman2008quantifying}. It performs a different grid search, across possible thresholds for converting continuous classifier outputs to binary positive positive or negative labels. For each possible threshold, the red class-conditional density determines the ratio of red above and below the threshold, and similarly for blue. Together with the cumulative classifier density above and below the threshold, that uniquely determines the blue vs. red fraction below the threshold and also the fraction above the threshold (Q). And that, in turns, determines the total blue and red areas, implying a target estimate (R). There is no strong reason to prefer any one threshold over another; this technique resolves that by taking the median of all the estimates yielded by different choices of classifier threshold. This lacks theoretical justification, but this threshold varying approach worked well in some empirical datasets~\cite{forman2008quantifying,gonzalez2017review}.

\section{Impacts of Alternative Design Elements on the Accuracy of Prevalence Estimates}
\label{sec:impacts}

This section uses simulated data to explore the impacts of different design elements in the prevalence estimation process on the accuracy of estimates. The analysis here yields insights about the impact of classifier predictive power. It also yields intuitions about the errors that will occur when there is a mismatch between the true data generating process and the stability assumption that is made.

Our procedure is to simulate base and target datasets from joint distributions with known parameters, run several alternative estimation processes, and compare the resulting estimates with the correct prevalence values that can be read off from the original parameters. We generate large enough datasets so that sampling variance is small; differences in the prevalence estimates are caused by alternative stability assumptions modeled into the estimation processes.

\subsection{Dataset and Classifier Simulation Process}

We use two data generating approaches for our simulations, because each provides scenarios that yield different insights. A joint distribution between the ground truth label $y$ and classifier output $C(\bx)$ is sufficient to simulate a dataset in which every item there has a ground truth label and a classifier score. Recall from~\Cref{fig:prevalence_framework} that a joint distribution can be specified either by the combination of a classifier score density and a calibration curve, or by a ground truth label prevalence and the two class-conditional densities. Our first data generator assumes stable class-conditional densities and uses two different label densities for the base and target datasets. Our second data generator assumes a stable calibration curve and uses two different classifier score densities, one for base dataset and one for target dataset.\footnote{Our data simulation and prevalence estimation tool is publicly available at \url{https://github.com/avalanchesiqi/pyquantifier}}

\subsubsection{Intrinsic data generator \& Stable class-conditional densities} 
As described in~\Cref{sec:background}, this is the intrinsic data generating process ($y \rightarrow \bx$). We model the joint distribution as a mixture of two Beta distributions, one for positive class $\mathcal{B}_\oplus(\alpha_\oplus, \beta_\oplus)$ and one for negative class $\mathcal{B}_\ominus(\alpha_\ominus, \beta_\ominus)$. We specify the label probability for a base dataset $P_B(y=\oplus)$ and for a target dataset $P_T(y=\oplus)$. To simulate an item, we first make a Bernoulli draw according to the label probability, yielding a ground truth label $y_i \in \{\oplus, \ominus\}$. We then generate a classifier score $C(\bx) \in [0, 1]$ as a draw from the corresponding Beta distribution.

Note that the four parameters that define the two Beta distributions are kept fixed between the base and target datasets. The only parameter that changes between $B$ and $T$ is the label probability. Thus, with this approach we can generate a dataset from the following five parameters:
$(\underbrace{\alpha_\oplus, \beta_\oplus, \alpha_\ominus, \beta_\ominus}_{\textbf{class-conditional densities}}, \underbrace{P(y = \oplus)}_{\textit{label probability}})$. The invariant properties are \textbf{bolded}, while the varying property is \textit{italicized}.

In our simulations using this approach, we always use 20\% positive for the base dataset, $P_B(y = \oplus)=0.2$, and 60\% positive for the target dataset, $P_T(y = \oplus)=0.6$. For the two Beta distributions, we design two scenarios: 

\begin{itemize}
    \item strong classifier: $\mathcal{B}_\oplus(10, 2)$, $\mathcal{B}_\ominus(2, 5)$
    \item weak classifier: $\mathcal{B}_\oplus(7, 6)$, $\mathcal{B}_\ominus(2, 5)$
\end{itemize} 

These two intrinsic data generators are illustrated in the first plots in rows 1 and 2 of both \Cref{table:calib_experiments} and \Cref{table:extra_experiments}. Note that we have superimposed the two class-conditional density functions. This is clear in row 1, where there is little overlap, indicating that the classifier has high predictive power; while the substantial overlap in row 2 shows a weak classifier that does not discriminate well between positives and negatives.

\subsubsection{Extrinsic data generator \& Stable calibration curve}
As described in the~\Cref{sec:background}, this is the extrinsic data generating process ($\bx \rightarrow y$), but requires a somewhat stronger assumption, that the calibration curve $P(y|C(\bx))$ is the same for the target dataset as for the base dataset. In this approach, we simulate an item as a draw from a classifier score density function, which we model as a mixture of two Beta distributions with five parameters, $\lambda \mathcal{B}_1(\alpha_1, \beta_1) + (1 - \lambda) \mathcal{B}_2(\alpha_2, \beta_2)$. This yields a realized classifier score, $C(\bx_i) \in [0, 1]$. To generate a ground truth label $y_i$, we first find the calibrated probability associated with that classifier score, using a pre-defined calibration curve $Calib(\cdot)$. We then draw the ground truth label $y_i$ as a Bernoulli draw with the computed calibrated probability $Calib(C(\bx_i))$. Our calibration curve is specified by two parameters, $w$ and $b$, as would result from a logistic regression, i.e., formula of Platt scaling~\cite{platt1999probabilistic}:
\begin{equation*}
Calib(w, b) = \frac{\mathrm{1}}{\mathrm{1} + e^{w \cdot C(\bx) + b}} \\
\label{eq:calibration_curve}
\end{equation*}

We use the same calibration curve, with fixed parameters $w$ and $b$, for both the base and target datasets, but vary the classifier score density curve. Thus, we have a total of seven parameters to generate a dataset in an extrinsic fashion, $(\underbrace{w, b}_{\textbf{calibration curve}}, \underbrace{\alpha_1, \beta_1, \alpha_2, \beta_2, \lambda}_{\textit{classifier score density}})$. The invariant properties are \textbf{bolded}, while the varying properties are \textit{italicized}.

In our simulations using this approach, we design the following classifier density score functions: 
\begin{itemize}
    \item base dataset: $C(\bx) \sim 0.2 \mathcal{B}_1(10, 2) + 0.8 \mathcal{B}_2(2, 5)$
    \item target dataset: $C(\bx) \sim 0.6 \mathcal{B}_1(10, 2) + 0.4 \mathcal{B}_2(2, 5)$
\end{itemize}
These base and target classifier score densities are the second and third plots in rows 3 and 4 of~\Cref{table:extra_experiments}. We simulate a strong and a weak classifier, and keep their parameters fixed between the base and target datasets:
\begin{itemize}
    \item strong classifier: $Calib(w=25, b=-15)$
    \item weak classifier: $Calib(w=0.5, b=-1)$
\end{itemize}

\begin{table*}[ht]
    \centering
    \small
    \begin{tabular}{c|c|c|c|c|c}
        \toprule
        Id & Generating properties of & Estimated properties of & \multicolumn{2}{c}{Prevalence estimator} & True base \\
        & base joint distribution & base joint distribution & calib. probabilistic est. & no calib. & prevalence\\
        \midrule
        1-intrinsic-strong & {{\begin{minipage}{.15\textwidth}
        \includegraphics[width=\linewidth]{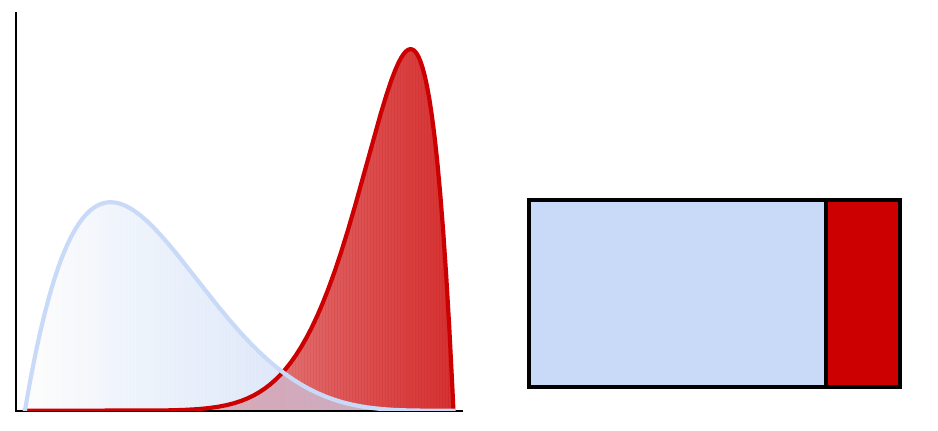}
        \end{minipage}}} & {\begin{minipage}{.22\textwidth}
        \includegraphics[width=\linewidth]{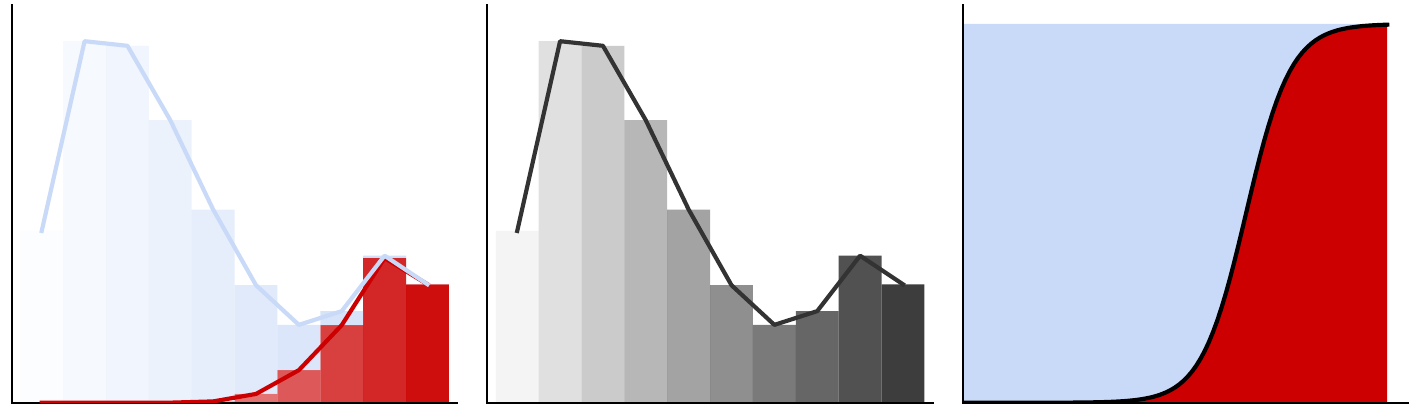}
        \end{minipage}} & \textbf{\begin{tabular}{@{}c@{}}20.40\% \\ (19.86\%, 20.91\%)\end{tabular}} & 39.68\% & 20.00\% \\
        2-intrinsic-weak & {{\begin{minipage}{.15\textwidth}
        \includegraphics[width=\linewidth]{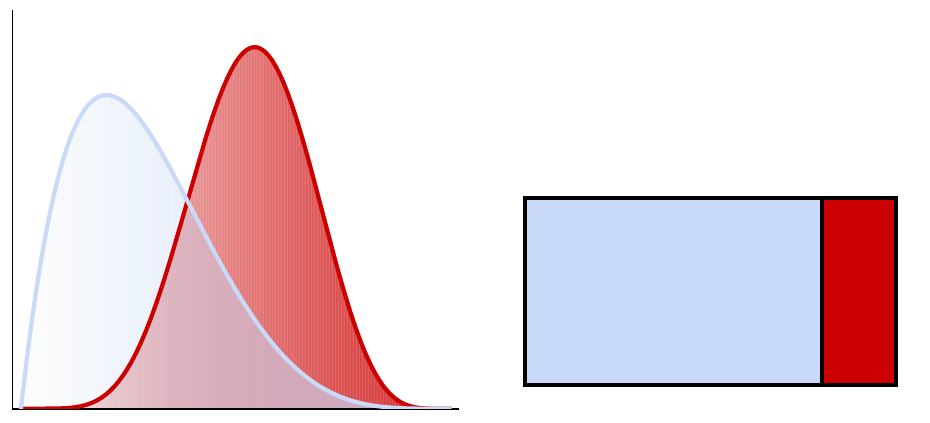}
        \end{minipage}}} & {\begin{minipage}{.22\textwidth}
        \includegraphics[width=\linewidth]{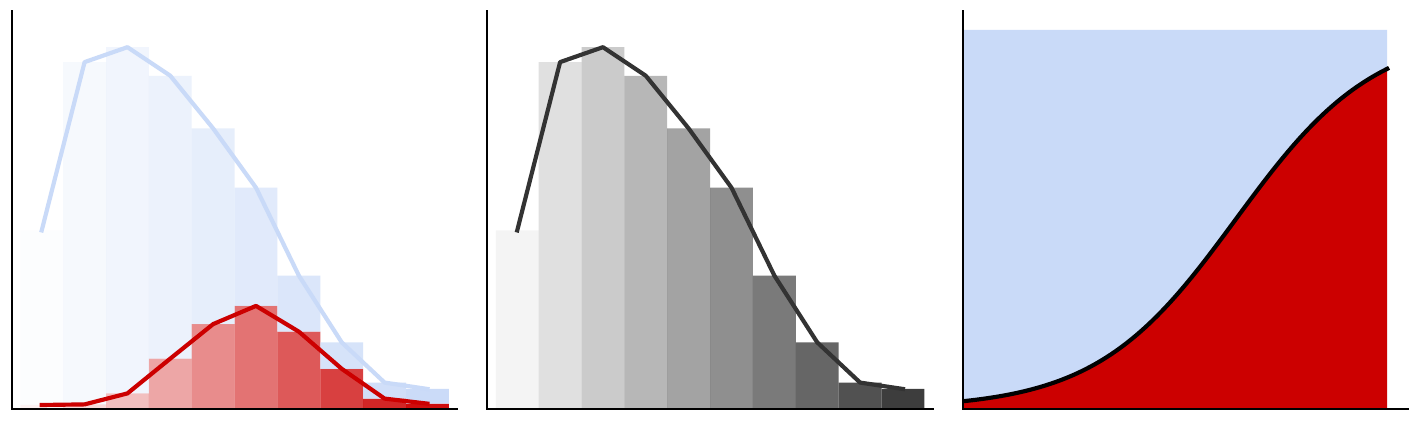}
        \end{minipage}} & \textbf{\begin{tabular}{@{}c@{}}19.71\% \\ (18.28\%, 21.21\%)\end{tabular}} & 35.36\% & 20.00\% \\
        \midrule
        3-extrinsic-strong & {{\begin{minipage}{.15\textwidth}
        \includegraphics[width=\linewidth]{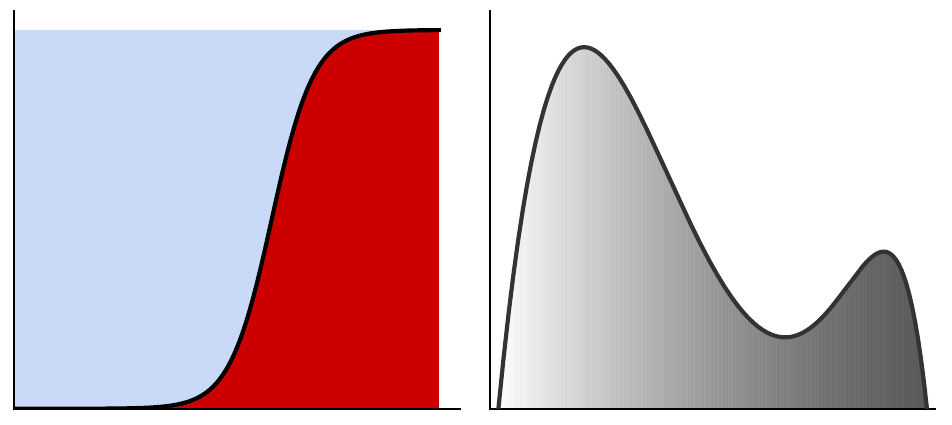}
        \end{minipage}}} & {\begin{minipage}{.22\textwidth}
        \includegraphics[width=\linewidth]{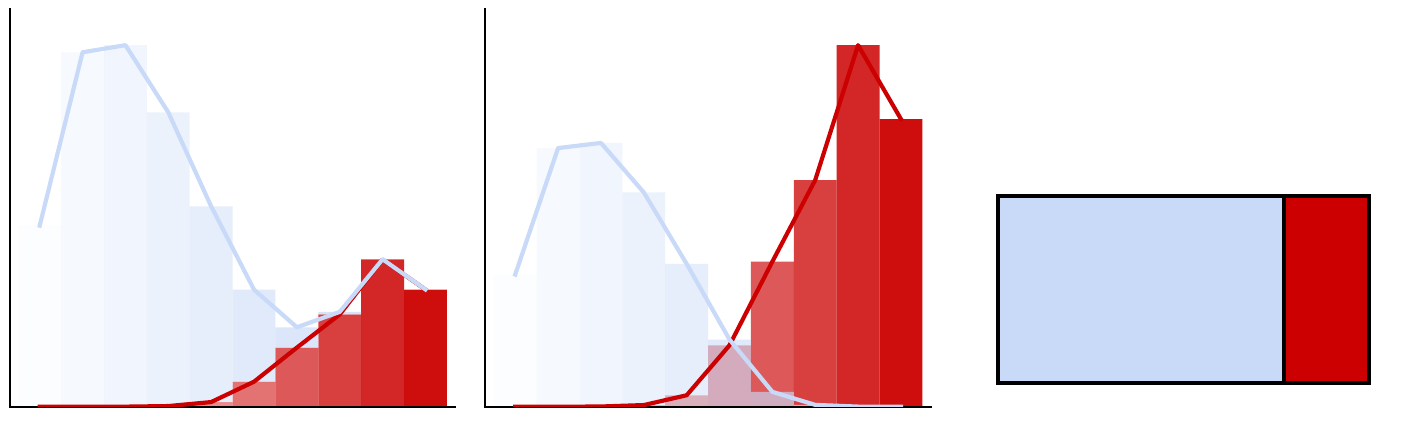}
        \end{minipage}} & \textbf{\begin{tabular}{@{}c@{}}24.14\% \\ (23.55\%, 24.67\%)\end{tabular}} & 39.41\% & 23.13\% \\
        4-extrinsic-weak & {{\begin{minipage}{.15\textwidth}
        \includegraphics[width=\linewidth]{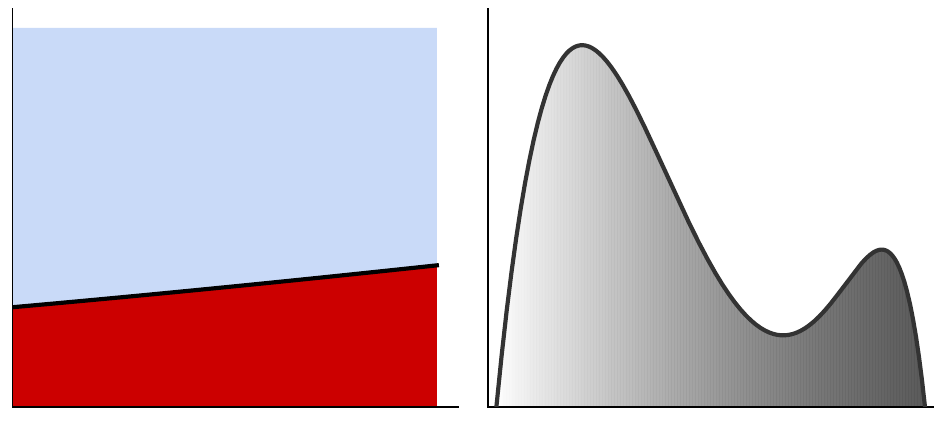}
        \end{minipage}}} & {\begin{minipage}{.22\textwidth}
        \includegraphics[width=\linewidth]{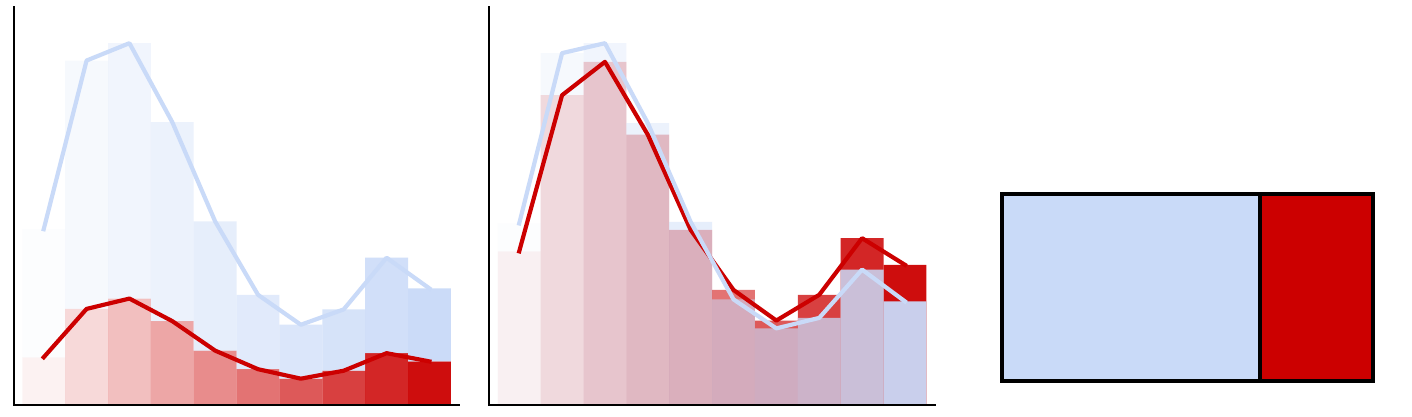}
        \end{minipage}} & \textbf{\begin{tabular}{@{}c@{}}31.25\% \\ (28.89\%, 33.17\%)\end{tabular}} & 39.45\% & 31.22\% \\
        \bottomrule
    \end{tabular}
    \caption{Calibration phase experiment results. For experiment group 1-4, we specify two properties (first column plots) that are sufficient to define a joint distribution between $C(\bx)$ and $y$. We can infer other properties of the same base joint distribution and visualize them in the second column plots. Numbers in the bracket indicate 95\% confidence intervals via bootstrapping.
    }
    \label{table:calib_experiments}
\end{table*}

\begin{table*}[hbt!]
    \centering
    \small
    \begin{tabular}{c|c|c|c|c|c|c}
        \toprule       
        Id & Stable & Non-stable & Inferred & \multicolumn{2}{c|}{Prevalence estimator} & True target \\
        & generating & generating properties & properties & mixture model & calib. probabilistic est. & prevalence\\
        & properties & ($B \rightarrow T$) & ($B \rightarrow T$) & & & \\
        \midrule
        & \multirow{2}{*}{{\begin{minipage}{.06\textwidth}
        \includegraphics[width=\linewidth]{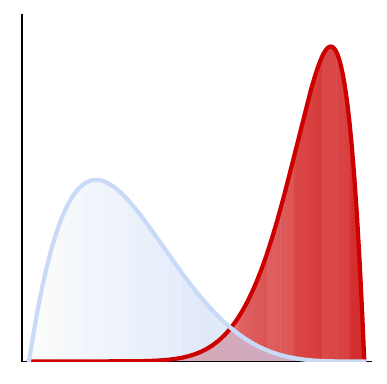}
        \end{minipage}}} & \multirow{2}{*}{{\begin{minipage}{.06\textwidth}
        \includegraphics[width=\linewidth]{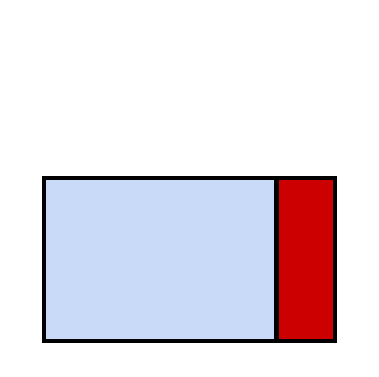}
        \end{minipage}} $\rightarrow$ {\begin{minipage}{.06\textwidth}
        \includegraphics[width=\linewidth]{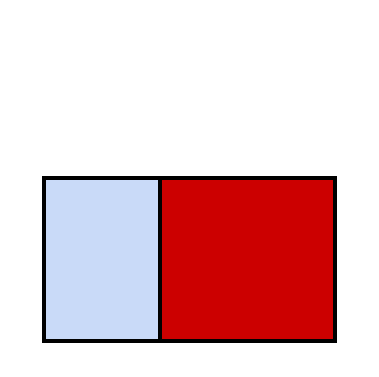}
        \end{minipage}}} & \multirow{2}{*}{{\begin{minipage}{.06\textwidth}
        \includegraphics[width=\linewidth]{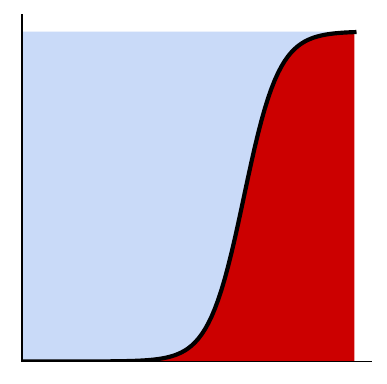}
        \end{minipage}} $\rightarrow$ {\begin{minipage}{.06\textwidth}
        \includegraphics[width=\linewidth]{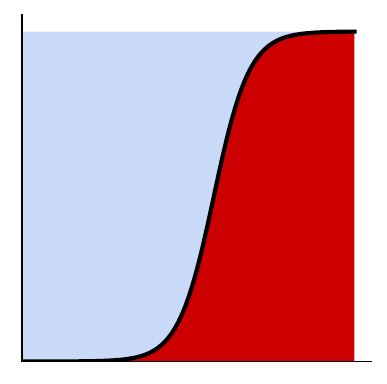}
        \end{minipage}}} & & \\
        \begin{tabular}{@{}c@{}}1-intrinsic\\-strong\end{tabular} & & & & \textbf{\begin{tabular}{@{}c@{}}60.21\% \\ (60.00\%, 61.00\%)\end{tabular}} & \begin{tabular}{@{}c@{}}\underline{52.63\%} \\ (51.60\%, 53.30\%)\end{tabular} & 60.00\% \\
        & \multirow{2}{*}{{\begin{minipage}{.06\textwidth}
        \includegraphics[width=\linewidth]{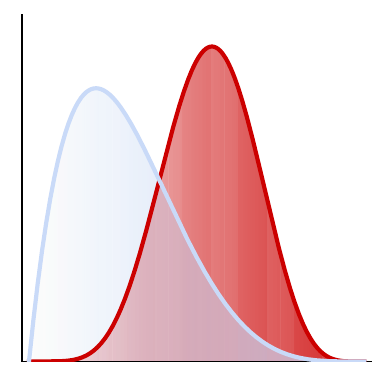}
        \end{minipage}}} & \multirow{2}{*}{{\begin{minipage}{.06\textwidth}
        \includegraphics[width=\linewidth]{image-files/table2/table2_row1_shift_b.pdf}
        \end{minipage}} $\rightarrow$ {\begin{minipage}{.06\textwidth}
        \includegraphics[width=\linewidth]{image-files/table2/table2_row1_shift_t.pdf}
        \end{minipage}}} & \multirow{2}{*}{{\begin{minipage}{.06\textwidth}
        \includegraphics[width=\linewidth]{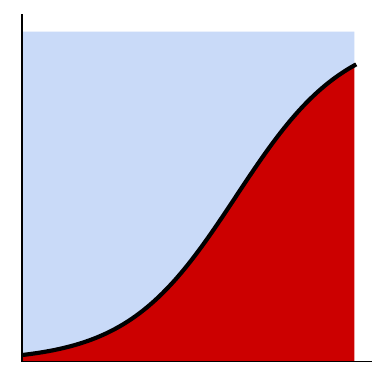}
        \end{minipage}} $\rightarrow$ {\begin{minipage}{.06\textwidth}
        \includegraphics[width=\linewidth]{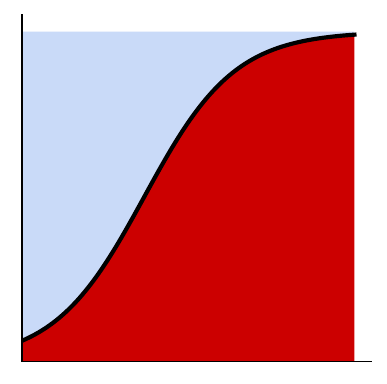}
        \end{minipage}}} & & & \\
        \begin{tabular}{@{}c@{}}2-intrinsic\\-weak\end{tabular} & & & & \textbf{\begin{tabular}{@{}c@{}}60.51\% \\ (57.00\%, 63.00\%)\end{tabular}} & \begin{tabular}{@{}c@{}}23.91\% \\ (22.57\%, 25.72\%)\end{tabular} & 60.00\% \\
        \midrule
        & \multirow{2}{*}{{\begin{minipage}{.06\textwidth}
        \includegraphics[width=\linewidth]{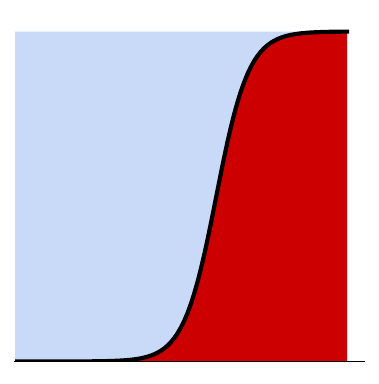}
        \end{minipage}}} & \multirow{2}{*}{{\begin{minipage}{.06\textwidth}
        \includegraphics[width=\linewidth]{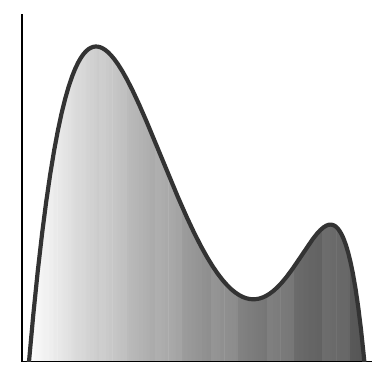}
        \end{minipage}} $\rightarrow$ {\begin{minipage}{.06\textwidth}
        \includegraphics[width=\linewidth]{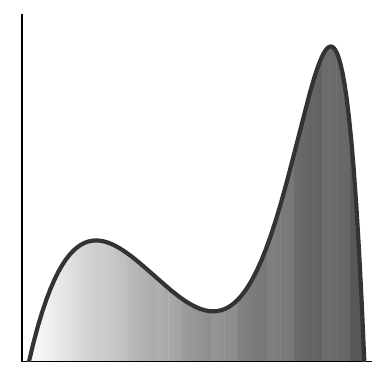}
        \end{minipage}}} & \multirow{2}{*}{{\begin{minipage}{.06\textwidth}
        \includegraphics[width=\linewidth]{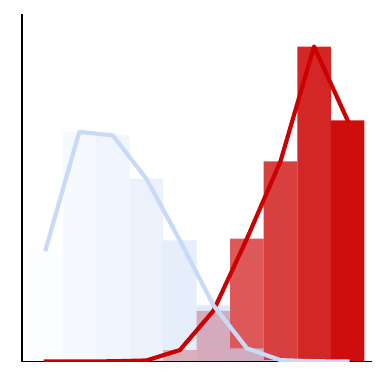}
        \end{minipage}} $\rightarrow$ {\begin{minipage}{.06\textwidth}
        \includegraphics[width=\linewidth]{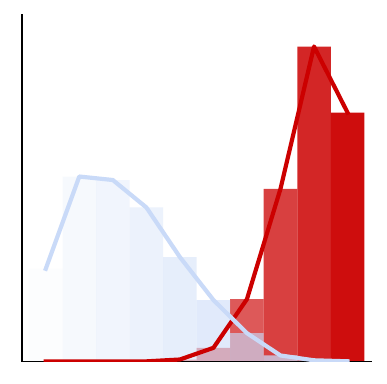}
        \end{minipage}}} & & & \\
        \begin{tabular}{@{}c@{}}3-extrinsic\\-strong\end{tabular} & & & & \begin{tabular}{@{}c@{}}\underline{63.11}\% \\ (62.48\%, 64.00\%)\end{tabular} & \textbf{\begin{tabular}{@{}c@{}}58.16\% \\ (57.61\%, 58.69\%)\end{tabular}} & 59.72\% \\
        & \multirow{2}{*}{{\begin{minipage}{.06\textwidth}
        \includegraphics[width=\linewidth]{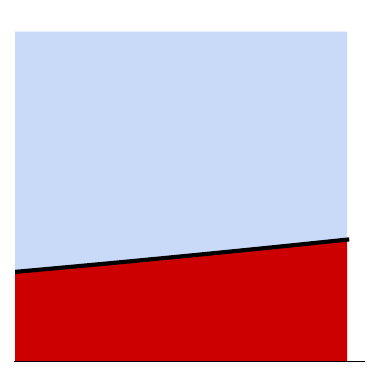}
        \end{minipage}}} & \multirow{2}{*}{{\begin{minipage}{.06\textwidth}
        \includegraphics[width=\linewidth]{image-files/table2/table2_row3_shift_b.pdf}
        \end{minipage}} $\rightarrow$ {\begin{minipage}{.06\textwidth}
        \includegraphics[width=\linewidth]{image-files/table2/table2_row3_shift_t.pdf}
        \end{minipage}}} & \multirow{2}{*}{{\begin{minipage}{.06\textwidth}
        \includegraphics[width=\linewidth]{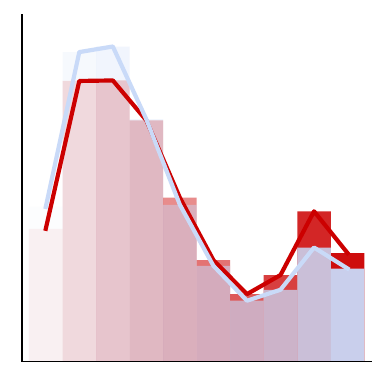}
        \end{minipage}} $\rightarrow$ {\begin{minipage}{.06\textwidth}
        \includegraphics[width=\linewidth]{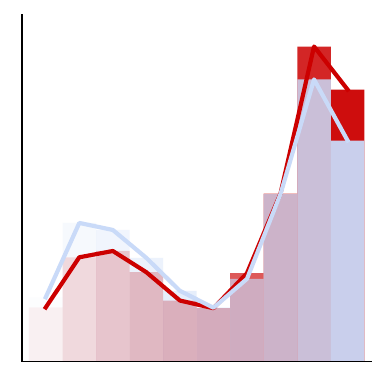}
        \end{minipage}}} & & & \\
        \begin{tabular}{@{}c@{}}4-extrinsic\\-weak\end{tabular} & & & & \begin{tabular}{@{}c@{}}97.97\% \\ (91.75\%, 100.00\%)\end{tabular} & \textbf{\begin{tabular}{@{}c@{}}33.32\% \\ (31.19\%, 35.51\%)\end{tabular}} & 33.38\% \\
        \bottomrule
    \end{tabular}
    \caption{Extrapolation phase experiment results. We keep one property fixed between the base and target datasets (first column plots). We vary the non-stable generating property (second column plots). We could make an alternative stability assumption and we show the inferred property needed for the corresponding estimation technique (third column plots). Numbers in the bracket: 95\% confidence intervals. Bolded numbers: estimates under the correct stability assumption. 
    }
    \label{table:extra_experiments}
\end{table*}

These two intrinsic data generators are illustrated in the first plots in rows 3 and 4 of both \Cref{table:calib_experiments} and \Cref{table:extra_experiments}. The strong classifier has a steep slope with a calibrated probability near 0 or 1 for most classifier scores. On the other hand, the weak classifier has similar calibrated probabilities for low and high scores, suggesting little discriminating power. 

For each combination of (intrinsic or extrinsic) data generator and (strong or weak) classifier, we generate a 20,000-item base dataset and a 20,000-item target dataset. Then, we apply the Calibrate-Extrapolate framework as described in~\Cref{fig:prevalence_framework}. For the calibration phase, we curate a calibration sample of up to 2,000 items by sampling up to 200 from each of ten classifier score strata (0-10\%; 10-20\%; etc.), less if there are fewer than 200 items in the respective bin. We use the Platt scaling method to estimate the calibration curve. For the extrapolation phase, we make prevalence estimates using two alternative techniques: mixture model and probabilistic estimator. To construct confidence intervals, we repeat the entire process of calibration and extrapolation 1,000 times, taking different calibration samples each time. We report the mean of the bootstrapped estimates and treat the middle 95\% as the confidence intervals, following the percentile bootstrap procedure~\cite{diciccio1988review}.

\subsection{Impacts of Classifier Predictive Power in the Calibration Phase}

\Cref{table:calib_experiments} shows the experiment results for the calibration phase for four groups under different data generating processes and different classifier predictive powers. We find that, from calibration sample to base dataset, one can and should assume a stable calibration curve because the sampling is conducted based on $C(\bx)$. Taking the example of Neyman allocation, the sampling within each stratum is still random; but over the entire span of $C(\bx)$, it is not. The class-conditional densities would thus change from base dataset to calibration sample, invaliding prevalence estimation techniques relying on that assumption. Passing the raw classifier scores to a calibration curve learned from the annotated sample, and then summing up the calibrated probabilities can accurately recover the ground truth prevalence, regardless of the data generator or classifier power (\textbf{bolded numbers}). Simply summing up the uncalibrated classifier scores (no calib. columns) yields bad estimation performance. Nevertheless, we remark that weak classifiers still have wider confidence intervals in the repeated measures.

\subsection{Sensitivity to Violations of Stability Assumptions in the Extrapolation Phase}

\Cref{table:extra_experiments} shows the experiment results for the extrapolation phase for four pairs of (base $\rightarrow$ target) comparisons. We find that, from base dataset to target dataset, one has to choose the stability assumption compatible with the data generating process to make a good prevalence estimate. That is, mixture model works well for intrinsic data generator and calibrated probabilistic estimator for extrinsic data generator. If the stable property between base and target dataset is correctly picked, the estimates will be accurate, regardless of classifier power (\textbf{bolded numbers}). Second, even under a wrong stability assumption, a strong classifier can still make decent estimates, as shown by the underlined numbers. This clarifies the value of training a high quality classifier, for its robustness to the violations of stability assumptions in prevalence estimation.

\subsection{Discussion}

We remark on a crucial difference between the sampling approach in our framework and in previous quantification literature~\cite{forman2005counting,gonzalez2017review}. In those papers, the authors started with the ground truth labels $y$, and artificially created test datasets with different levels of prevalence conditioned on $y$. For example, quoting the experiment protocol from~\cite{forman2005counting}, \textit{``we randomly drew 200 positives and 1000 negatives from each benchmark classification task as the maximum training set... we randomly removed positives or negatives to achieve various desired testing class distributions''}.

Sampling based on the ground truth is attractive, because it would preserve the class-conditional density function for each class, thus enabling estimation techniques that rely on the stable class-conditional density assumption. However, sampling based on ground truth $y$ is less realistic in social media research because it is difficult to obtain the ground truth for every item. Instead, CSS researchers are usually more interested in making unbiased prevalence estimates for large-scale social data and monitoring the trend of prevalence. In addition, they usually have access to high quality pre-trained models. They can use these models to strategically select a better sample for ground truth labels, for instance, oversampling rare classes. This can improve the performances in both classification and prevalence estimation models built from the calibration sample.

In addition, assuming only a stable calibration curve between the base and target datasets limits the range of possible final estimates. For example, if the learned calibration curve for base dataset $P_B(y|C(\bx))$ is bounded between $20\%$ and $60\%$, it means that the estimated prevalence for any target datasets would never exceed $60\%$, nor fall short of $20\%$. On the other hand, assuming only class-conditional densities amplifies observed changes; it can lead to estimates clipped at 0\% or 100\%. For example, if adding any amount of negative class density to the positive class density makes the mixed classifier score density less similar to the observed classifier score density in a target dataset, the estimator would output 100\%.

\section{Application: How Many Toxic Comments Are Posted on Social Media Everyday?}
\label{sec:application}

We apply the Calibrate-Extrapolate framework to a real world application that estimates the extent to which online comments on each week's news stories would be perceived as toxic by the general public. Separate estimates are made for three platforms: Reddit, Twitter/X, and YouTube. For each, we collected human labels for one calibration sample, then conducted an extrapolation phase for each week for one year. We performed two estimations assuming either stable calibration curves (calibrated probabilistic estimator) or stable class-conditional densities (mixture model), in order to illustrate sensitivity to different stability assumptions.

\subsection{Public News Comments Dataset}

Details of the data collection are described in a separate report.\footnote{\url{https://csmr.umich.edu/projects/hot-speech/}} We report a summary here. The contribution of this paper is to explore techniques for estimating prevalence in datasets like this one; the dataset itself is not a contribution. 

Each day, the 1,000 most engaging URLs on Facebook and on Twitter/X were collected, as reported by NewsWhip.\footnote{\url{https://www.newswhip.com/}} Following the semi-supervised approach used in~\cite{bakshy2015exposure}, a machine learning classifier was trained to identify the URLs that reference ``hard news'' (e.g., politics, economics) as opposed to softer news (e.g., sports, music). For each hard news URL, a search was conducted for posts on Reddit and Twitter/X that contain the URL and for YouTube videos whose titles match the news headlines. All comments on the matched posts and videos were scraped, excluding deleted, moderated, and bot comments on Reddit; retweets and quoted tweets on Twitter/X.

To make the final metrics comparable across platforms, we selected an equal number of comments from each platform about each hard news URL -- the minimum number available across the three platforms. This sampling strategy avoids the risk that controversial stories, whose comments are frequently toxic, might garner more comments on one platform than on another. We did this to ensure that differences in the measured frequency of toxic comments reflect different modes of expression about the same stories rather than differences of popularity across platforms. 

We report on target datasets of comments on URLs from 2022. There were 15,564 hard news URLs having at least ten comments on all three platforms, about 42.6 per day. On average the collection includes 5,631 distinct comments per day on Reddit, 5,355 on Twitter/X, and 4,271 on YouTube. The daily comment numbers are different over three platforms because some YouTube comments were selected for multiple URLs with similar headlines.

The comments posted during August, 2021 form our base dataset. A disproportionate random sampling process was conducted, loosely based on Neyman allocation, yielding 1,144 Reddit comments, 1,154 Twitter/X replies, and 1,162 YouTube comments. These were annotated by Amazon Mechanical Turk (MTurk) workers, who were asked to label whether a comment was toxic, with the same definition used when collecting training data for the Perspective API~\cite{wulczyn2017ex}---\textit{``a rude, disrespectful, or unreasonable comment that is likely to make readers want to leave a discussion''}. Five workers labeled each comment; the ground truth label is the majority vote of the five. 

The annotation was carried out in November, 2021. We limited access to U.S. residents who had completed at least 1,000 Human Intelligence Tasks (HITs) with $\geq 98\%$ acceptance rate. They first had to complete a qualification task, in which we provided the concept definitions, annotation instructions, five practice comments with labeling answers given by domain experts, and four questions with clear answers. To qualify, a worker had to answer all four questions correctly. Qualified workers were then permitted to label up to 100 comments, in 10-comment batches. HITs were priced with a target hourly pay rate of \$15.

\subsection{Prevalence Estimates Using Perspective API}

For the black-box classifier, we used the Perspective API.\footnote{\url{https://perspectiveapi.com/}} We detected that the Perspective API changed significantly in May 2022. For many comments, the Perspective API returned a much lower score after that date. Thus, we can think of there being two classifiers, both called Perspective API. 

\begin{figure}[tbp]
    \centering
    \includegraphics[width=1\linewidth]{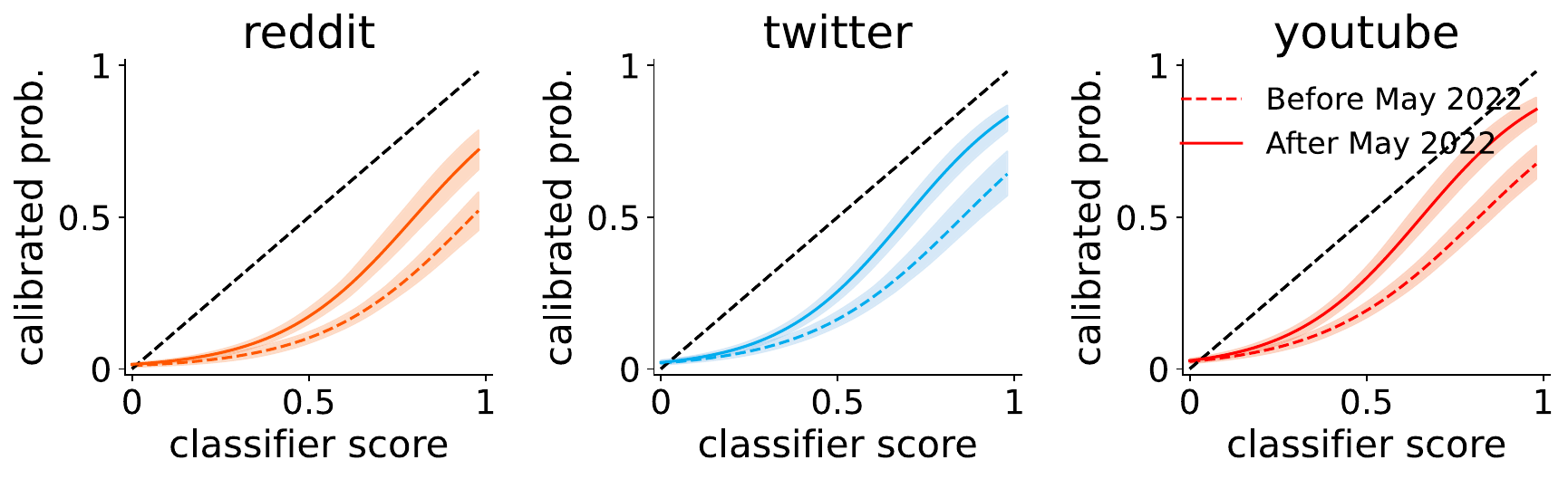}\\
	\caption{Calibration curves for two versions of Perspective API, both against the calibration sample from August 2021.
	}
	\label{fig:toxic_calib_curves}
\end{figure}

\Cref{fig:toxic_calib_curves} shows the calibration curves for the two classifiers, estimated from the labeled calibration sample, using the Platt scaling technique. Since scores were generally lower with the later classifier, the calibrated probability associated with any given score was higher after May 2022. Note that in both periods, however, the calibration curves lie entirely below the 45 degree line. This indicates that Perspective API scores are systematically higher than the true calibrated probability in our base dataset: for example, on comments where the Perspective API outputs 0.5, less than half of comments were labeled as toxic by MTurk workers. 

Also note that the calibration curves, while similar, are not exactly the same for the three platforms. The calibrated probability of being labeled toxic was higher for a comment from YouTube receiving a Perspective API score of 0.5 than for a comment from the other platforms with the same score. This underscores the need to calibrate a classifier separately when using it on datasets that have distinct error profiles.

\begin{figure}[tbp]
    \centering
    \includegraphics[width=0.99\linewidth]{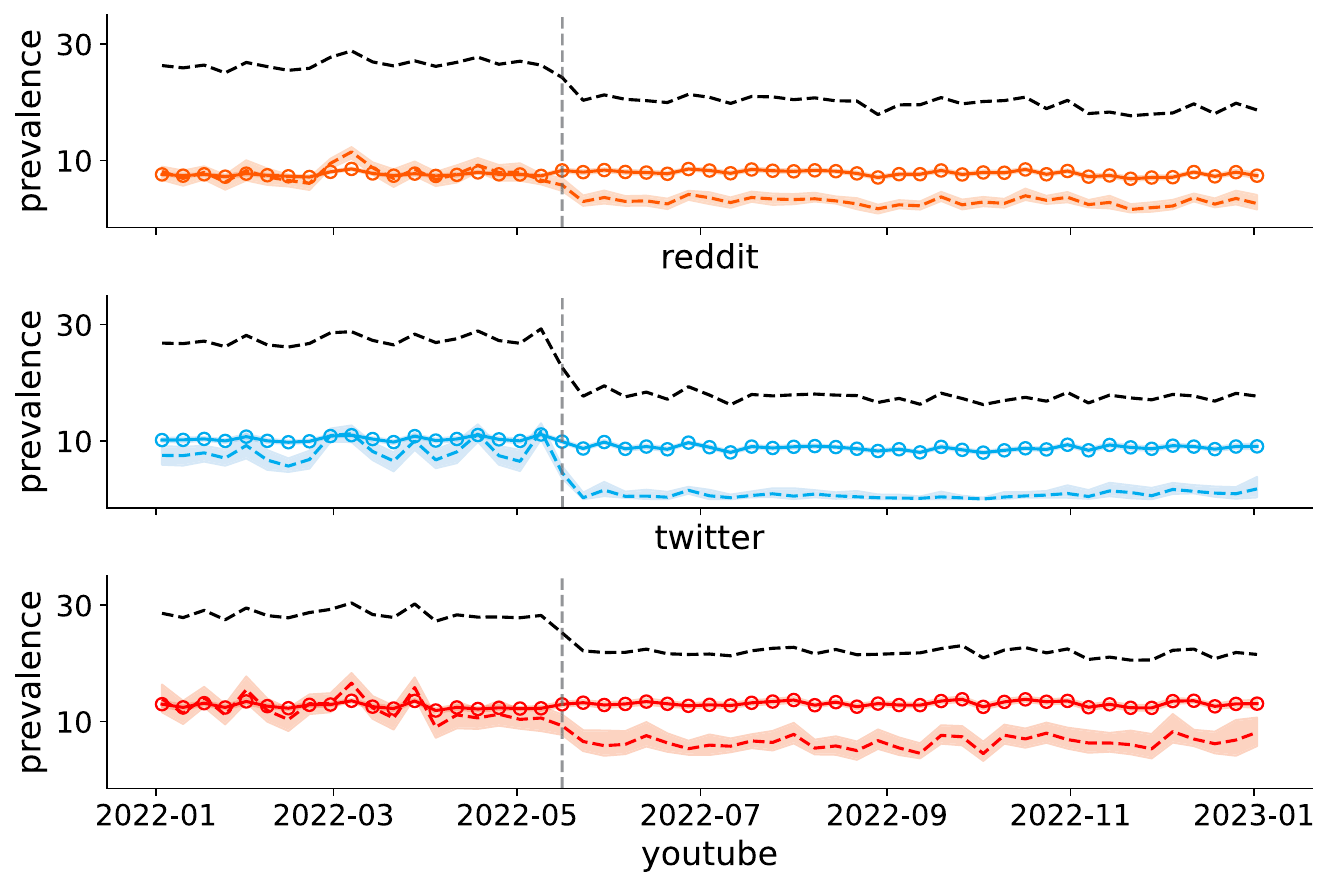}\\
    \caption{Toxicity prevalence estimates using three alternative estimation techniques. Colored circled: assumed stable calibration curve; colored dashed: assumed stable class-conditional densities; black dashed: without calibration. 
    }
    \label{fig:toxic_prevalence}
\end{figure}

\Cref{fig:toxic_prevalence} shows our estimates of perceived toxic comments for each week in 2022, for each of the three platforms. The black dashed line uses the Perspective API scores as if they were calibrated probabilities, effectively following the unadjusted ``probabilistic classify and count'' approach. The vertical dashed line pinpoints when the classifier changed, yielding lower prevalence estimates. However, since classifier scores still were higher than their calibrated values, the estimates without calibration are consistently too high for all three platforms, by very large margins.

The two colored lines for each platform show the results of the extrapolation phase estimation assuming either a stable calibration curve from base to target dataset (calibrated probabilistic estimator technique) or stable class-conditional densities (mixture model approach). Given the sizes of calibration sample and each week's datasets, the confidence intervals due to sampling error are fairly tight. However, the difference in estimates using the two techniques, based on different stability assumptions, yields much larger differences. Assuming stable class-conditional densities leads to more extreme estimates: when the target period scores are lower than the base period, the prevalence estimate moves lower by even more. This can even lead to different conclusions about the ordering of platforms. For example, assuming stable class-conditional densities leads to a conclusion that Twitter/X had lower toxicity than Reddit towards the end of 2022, but assuming a stable calibration curve leads to a conclusion that Twitter/X had higher toxicity than Reddit. 

Note that the calibrated probabilistic estimator technique seems to have handled the May changeover in the Perspective API better than the other two approaches. While the other approaches show sharply lower prevalence estimates as soon as the Perspective API gave lower scores, with the calibrated probabilistic estimator, the estimated prevalence was slightly higher on Reddit and YouTube in the weeks after the change but slightly lower on Twitter/X. Since it is unlikely that there was a major and long-lived change in the actual toxicity on the platforms immediately following the Perspective API change, the fact that the probabilistic estimator technique gives similar results using either version of Perspective API is one reason to think that its assumptions are better suited to this toxicity prevalence estimation task. 


\section{Conclusion}
\label{sec:conclusion}

In this work, we present the ``Calibrate-Extrapolate'' framework. It rethinks the prevalence estimation process as calibrating the classifier outputs against ground truth labels to obtain the joint distribution of a base dataset and then extrapolating to that of a target dataset. Visualizing the joint distributions makes the stability assumption needed for a prevalence estimation technique clear and easy to understand. 

In the calibration phase, all techniques require only a stable calibration curve between a labeled calibration sample and the full base dataset. Disproportionate sampling based on the classifier outputs does not lead to a violation of that requirement and can improve the efficiency of prevalence estimation for a base dataset, providing tighter confidence intervals for estimates based on a limited number of labeled items. Beyond that efficiency improvement, however, a black-box classifier provides limited additional value in making a prevalence estimate for a single base dataset.

The real value of the black box classifier emerges in the extrapolation phase, where the classifier is applied to a target dataset but no further ground truth labels are collected. All prevalence estimation techniques for an extrapolation phase with no new labels depend on borrowing some property from the base dataset joint distribution, effectively assuming stability of that property from the base to the target dataset. Some techniques assume a stable calibration curve while some assume stable class-conditional densities. 

This framework provides a recipe for generating simulated datasets by specifying base and target joint distributions and using them as generators. By making instructive choices for these distributions, in \Cref{sec:impacts} we illustrated how estimation techniques can yield incorrect estimates when their stability assumptions are violated.

\Cref{sec:application} applied two of the techniques, as well as a benchmark that eschews calibration entirely, to a real world collection of 52 target datasets, one for each week in a year. There, the uncertainty due to sampling error was dwarfed by the uncertainty due to the choice of stability assumptions inherent in the prevalence estimation techniques.

\subsubsection{Practical advice for prevalence estimation}

We summarize a few key takeaways. A prevalence estimation tutorial developed by the authors can be found at this link.\footnote{\url{https://avalanchesiqi.github.io/prevalence-estimation-tutorial/}}

\begin{enumerate}
    \item It is never safe to make a prevalence estimate based on a classifier that was trained on a different dataset, without gathering some human labels for a calibration sample.
    \item If only a single prevalence estimate is needed, at a single point in time, or for a single dataset, a black-box classifier is of limited use.
    \begin{enumerate}
        \item If the ground truth labels are expected to be roughly balanced, the best approach is to select a random calibration sample and use the human label frequencies in the calibration sample as the prevalence estimate for the base dataset. 
        \item If the ground truth labels are expected to be unbalanced, use the classifier to produce a calibration sample with more balanced classifier scores. In this case, the rest of the calibration stage of the purposed framework should be followed. This will yield an unbiased estimate of the ground truth prevalence for the base dataset, with tighter confidence intervals.
        \item Converting continuous classifier outputs to binary labels based on a cutoff threshold (e.g., the classify and count approaches), will not generally yield reliable prevalence estimates.
    \end{enumerate}
    \item If prevalence estimates are needed for multiple related datasets, we suggest performing a calibration phase once for the base dataset and then applying the extrapolation phase to the remaining datasets.
    \begin{enumerate}
        \item For the calibration phase, use the classifier to select a sample and collect human labels and estimate the calibration curve, yielding an estimate of the full base dataset joint distribution. 
        \item For the extrapolation phase, reason about which stability assumption is more plausible. If it is an intrinsic data generation process, leading to stability of the class-conditional density functions, use the mixture model or median sweep. If the calibration curve from the base dataset can be assumed to also apply to the target dataset, use the probability estimator technique.
        \item One indicator about which stability assumption is better can come from a robustness check involving two independent classifiers. Using the same calibration dataset, estimate calibration curves and thus base joint distributions for both classifiers. Then, for each classifier, apply both approaches (e.g., distribution matching and probabilistic estimator) in the extrapolation phase. If using one of the techniques, both classifiers yield similar estimates for the target datasets, that technique's assumptions are more reasonable.
    \end{enumerate}    
\end{enumerate}

\subsubsection{Limitations and future directions}
The biggest limitation is that the work provides limited guidance on how to decide \textit{which} stability assumption is appropriate for particular real world scenarios. We provided some conceptual guidance, by connecting the stability assumptions to different data generating models in~\Cref{sec:background}. Some scenarios are more naturally described as intrinsic or extrinsic data generating processes. For example, some medical settings seem to have intrinsic data generation (a disease causes symptoms). Many social media settings, on the other hand, seem to have extrinsic data generation (the words writers choose cause readers to see messages as toxic or not). However, we point out that even if the same words are always perceived the same way, and classified the same way, that is not enough to ensure that the calibration curve for a classifier will be stable between base and target datasets. We provided a robustness approach: the assumption that yields similar prevalence estimates using different classifiers may be better.

In reality, there may be both prior probability shift and covariate shift between base and target datasets. And there can also be concept drift, where the ground truth labels for items change between datasets. A promising area for future research would be to develop guidance on how to diagnose when it is appropriate to take a classifier calibration curve estimated from one dataset and apply it to another. One approach would be to collect a second calibration sample from one of the target datasets, estimate another joint distribution, and see which properties were actually stable between the base joint distribution and that target joint distribution. Properties that were stable between those two might be reasonable to assume also stable for other target datasets that are constructed in a similar way.

Despite this limitation, the Calibrate-Extrapolate framework provides conceptual clarity about the logic behind alternative prevalence estimation techniques. Researchers and practitioners who think in terms of this framework will ask themselves the right questions about the techniques they would apply and can avoid many potential pitfalls that await the unwary prevalence estimator.

\section*{Acknowledgments}
This work was supported in part by the University of Michigan Center for Social Media Responsibility. We thank Nathan Teblunthuis, Libby Hemphill, Dallas Card, and Dirk Tasche for feedback that improved the paper and Chenxin Han for graphic design wizardry.

\bibliography{main}

\end{document}